\documentclass[twocolumn,aps,pra,longbibliography,superscriptaddress,nofootinbib]{revtex4-1}

\usepackage[utf8]{inputenc}
\usepackage{graphicx}
\usepackage{graphics}
\usepackage{amssymb}
\usepackage{color}
\usepackage{palatino}
\usepackage{nicefrac}
\usepackage{multibib}
\usepackage{bm}
\usepackage{mathrsfs}
\usepackage{amsmath}
\usepackage{amsfonts}
\usepackage{times,txfonts}
\usepackage{amsmath}
\usepackage{dsfont}
\usepackage{physics}

\begin{document}
\title{Analogue Penrose process in rotating acoustic black Hole}

\author{Decheng Ma}
%\email{madch90@gmail.com}
\affiliation{International Center of Quantum Artificial Intelligence for Science and Technology (QuArtist) \\
and Physics Department, Shanghai University, 200444 Shanghai, China}
\affiliation{Lanzhou Center for Theoretical Physics, and Key Laboratory of Theoretical Physics of Gansu Province, Lanzhou University, Lanzhou 730000, China}

\author{Enrique Solano}
\email{enr.solano@gmail.com}
\affiliation{International Center of Quantum Artificial Intelligence for Science and Technology (QuArtist) \\
and Physics Department, Shanghai University, 200444 Shanghai, China}
\affiliation{IKERBASQUE, Basque Foundation for Science, Plaza Euskadi, 5, 48009 Bilbao, Spain}
\affiliation{Kipu Quantum, Kurwenalstrasse 1, 80804 Munich, Germany}

\author{Chenglong Jia}
\affiliation{Key Laboratory for Magnetism and Magnetic Materials of the Ministry of Education, Lanzhou University, Lanzhou 730000, China}

\author{Lucas Chibebe C\'eleri}
\email{lucas@qpequi.com}
\affiliation{Institute of Physics, Federal University of Goi\'{a}s, 74001-970, Goi\^ania, Brazil}

\begin{abstract}
Analogue gravity stands today as an important tool for the investigation of gravitational phenomena that would be otherwise out of reach considering our technology. We consider here the analogue Penrose process in a rotating acoustic black hole based on a quantum fluid described by the draining bathtub model. Because of the rotating nature of this acoustic spacetime, a particle, that is a phonon travelling on the quantum fluid, will experience the well known frame dragging effect. By defining the effective Komar mass and angular momentum of this acoustic black hole, we found that energy can be extracted from it. At the same time, we show that the black hole angular momentum and mass are reduced, in complete analogy with the Penrose effect.
\end{abstract}

\maketitle

%%%%%%%%%%%%%%%%%%%%%%%%%%%%%%%%%%%%%%
%%%%%%%%%%%%%%%%%%%%%%%%%%%%%%%%%%%%%%
%%%%%%%%%%%%%%%%%%%%%%%%%%%%%%%%%%%%%%

\section{Introduction}

Analogue gravity can provide new insights for general relativity by investigating analogues of gravitational fields in controllable physical systems. The earliest analogue model of gravity dates back to 1923~\cite{Gordon1923,Felice1971}. However, it was due to W.~G.~Unruh‘s work on a formal analogy between sound waves in a moving fluid and light waves in a curved spacetime~\cite{Unruh1981} that initiated the modern era of gravitational analogues. Since then, a large number of systems has been proposed and explored for realizing analogue gravity, such as Bose-Einstein condensates~\cite{Weinfurtner2011,Torres2017}, slow light in optical systems~\cite{sheng2013,Tinguely2020,Bekenstein2017,Drori2019,Sheng2018,Zhong2018} and magnons in magnetic systems~\cite{Molina2017}. These analogue systems are not completely equivalent to general relativity, but they can capture many important features of real gravitational fields, which can be used as an important testing ground for general relativity, thus providing important insights on the quantum nature of gravity~\cite{Barcelo2011,Barcelo2019}.

Among the phenomena that can be studied in these systems, the Penrose process, a way to extract energy from rotating black holes~\cite{Penrose1971,Solnyshkov2019}, has become a new frontier. In order to simulate the Penrose process, an analogue of a rotating black hole is needed. It has been proposed that a dissipating vortex in a classical fluid could be used to simulate a rotating black hole~\cite{Visser1998}. Based on this idea, analogue rotating acoustic black holes have been demonstrated in classical~\cite{Torres2017} and quantum~\cite{Vocke2018} fluids. The related superradiance effect has also been theoretically~\cite{Basak2003,Patrick2018,Oliveira2010} and experimentally~\cite{Torres2017} investigated. Recently, with small vortices playing the role of massive particles moving in this acoustic spacetime, the analog Penrose process was studied in polariton condensates~\cite{Solnyshkov2019}. The basic idea is the creation of a vortex-antivortex pair in the ergosphere region. While the antivortex falls into the acoustic black hole, causing a reduction of its angular momentum, the vortex escapes from the acoustic black hole carrying more kinetic energy, thus mimicking the Penrose process. 

In this paper, we investigate the analogue Penrose process in a rotating acoustic black hole based on the draining bathtub model~\cite{Visser1998}. Although this model only describes an axisymmetric fluid with a constant density, it can exhibit an event horizon and an ergosphere, which can be used to simulate many interesting effects of rotating black holes in general relativity, including Hawking radiation and superradiance~\cite{Barcelo2011}. As we show in this paper, by using the local energy conservation and the existence of the ergoregion in the acoustic black hole, we can obtain the analogue Penrose process in this rotating acoustic black hole. In contrast to the specific and approximated model used in Ref.~\cite{Solnyshkov2019}, here we can demonstrate the general properties and behaviors of the acoustic analogue Penrose process.

Furthermore, we also resolve and clarify another important issue about the analogue Penrose process. In general relativity, during the Penrose process, energy is extracted form the rotating black hole that has its mass and angular momentum reduced. However, in the analogue gravity, because Einstein equation is absent, the acoustic spacetime is not directly linked with the energy-momentum of the underline system, and the mass and angular momentum of the effective acoustic black hole are not properly defined~\cite{Barcelo2011}. In order to give a quantitative description of the analogue Penrose process, we need to define the effective mass and angular momentum of the acoustic black hole in terms of the acoustic metric. Since the acoustic metric arising from the draining bathtub model is stationary and axisymmetric, by using the Komar integrals we can properly define the effective mass and angular momentum of the acoustic black hole, which are related to time translation and rotation symmetries of the acoustic metric, respectively. With these definitions, we can not only simulate the analogue Penrose process with draining bathtub model, like the results in Ref.~\cite{Solnyshkov2019}, but we are also able to analyse the variation of the effective mass and the angular momentum of the acoustic black hole during the process. Interesting, we also found an inequality between the amount of energy that can be extracted from the rotating acoustic black hole and the change in its angular momentum, analogously to what happens in the general relativistic system. However, the nature of both processes is very different, as mentioned in our final discussion.

The paper is organized as follows. Section~\ref{DBT metric} introduces the draining bathtub model as an analogue of a rotating black hole, as well as a discussion of the frame dragging effect in this acoustic spacetime. With the introduction of the appropriate Killing vectors, we define the surface gravity, the effective mass and angular momentum of the acoustic black hole. Based on these definitions, in Sec.\ref{Analogue Penrose} we then analyse the analogue Penrose process. Finally, we close the paper in Sec.~\ref{conclusion} with our final remarks.

%%%%%%%%%%%%%%%%%%%%%%%%%%%%%%%%%%%%%%%%%%%%%%%%%%%%%%%%
%%%%%%%%%%%%%%%%%%%%%%%%%%%%%%%%%%%%%%%%%%%%%%%%%%%%%%%%
\section{The acoustic metric}\label{DBT metric}

As we know, sounds waves in a moving fluid can be used to simulate the motion of light in a curved spacetime. The fluid can be classical, like Unruh's original proposal~\cite{Unruh1981}, or quantum, like a superfluid (eg.superfluid helium, Bose-Einstein condensate (BEC), etc.)~\cite{Barcelo2011}. Here, we consider a general incompressible fluid with constant density, so the speed of sound relative to the fluid is constant. We also assume that the viscosity of the fluid is zero, thus the internal friction of the fluid can be neglected. Such system can be realized by a superfluid helium or a superfluid BEC. Providing the flow of the fluid is locally irrotational, we can introduce a velocity potential $\theta_0$ to describe the motion of the fluid. The velocity of the flow is given by the gradient of the velocity potential $\mathbf{v}=-\nabla\theta_0$. When subjected to external disturbance, sound waves can be created in the fluid. Since the propagation of sound waves in a fluid is governed by a simple wave equation, then by regarding the sound waves as linear perturbations $\theta$ of the background velocity potential $\theta_0$, we can rewrite the wave equation of the sound waves into the form of a Klein-Gordon equation $\theta$~\cite{Barcelo2011,Visser1998,Oliveira2010},
\begin{equation}
    \frac{1}{\sqrt{- g}} \partial_{\mu} \left( \sqrt{- g} g^{\mu \nu}
   \partial_{\nu} \right) \theta = 0.
\end{equation}
As we can see, the sound waves in the fluid propagate like scalar fields on an effective spacetime described by the acoustic metric $g_{\mu\nu}$
\begin{equation}
  g_{\mu \nu} = \left[\begin{array}{ccc}
     - (c^2 - v^2) & \vdots & - v_j\\
     \cdots & \cdots & \cdots\\
     - v_i & \vdots & \delta_{i j}
   \end{array}\right],
\end{equation}
where $v_i$ is the $i$-th component of the flow velocity while $c$ stands for the speed of sound in the fluid.

As mentioned in the introduction, we need a rotating black hole if we want to investigate the Penrose effect. In order to create a rotating acoustic black hole that mimics the Kerr black hole, we use the draining bathtub model, which is built on a smoothly rotating and dissipative fluid flow with a sink at the origin\cite{ Visser1998}. In polar coordinates on the plane, the fluid velocity potential of the draining bathtub model is given by
\[ 
\theta_0 (r, \phi) =  A \log r - B \phi, 
\]
where $A$ and $B$ are real parameters. The corresponding fluid velocity profiled reads
\begin{equation}
  \mathbf{v}= - \frac{A}{r} \hat{r} + \frac{B}{r} \hat{\phi} .   
\end{equation}
The fluid flow of this model includes a radial and an irrotational component. The radial flow is created by the sink at the origin, its magnitude being determined  by the drain rate parameter $A$, which characterize the speed of the fluid to be drained out from the system. The irrotational flow arises from the circulation of the fluid, and its magnitude is determined by  the circulation parameter $B$.

Within the considered model, the effective spacetime experienced by the sound waves is defined by the $(2+1)$ dimensional acoustic metric 
\begin{equation}
    \begin{split}
 \dd s^2 =&  - \left( c^2 - \frac{A^2 + B^2}{r^2} \right)
  \dd t^2 + \frac{2 A}{r} \dd r \dd t - 2 B \dd \phi \dd t \\
  &+  \dd r^2 + r^2 \dd \phi^2 .
    \end{split}
\end{equation}
Now, by considering the coordinate transformation
\begin{equation*}
    \begin{split}
        \dd t& = \dd t^{\ast} + \frac{A r}{- A^2 + c^2 r^2} \dd r, \qquad \dd \phi = \dd
   \phi^{\ast} + \frac{A B}{r (- A^2 + c^2 r^2)} \dd r, \\ 
   r& = r^{\ast},
    \end{split}
\end{equation*}
the metric can be rewritten as
\begin{equation}\label{metric} 
\begin{split}
          \dd s^2 = &  - c^2 \left( 1 - \frac{A^2 + B^2}{c^2 r^2}
  \right) \dd t^2 + \left( 1 - \frac{A^2}{c^2 r^2} \right)^{- 1} \dd r^2 \\
    &- 2 B \dd \phi \dd t + r^2 \dd \phi^2 ,
\end{split}
\end{equation}
where we dropped the *-superscript for simplicity.

As we can see, once the radial component of the fluid
velocity exceeds the speed of sound, the $g_{rr}$ component of the metric vanishes, and  an acoustic event horizon forms at
\begin{equation}
  r_h = \frac{ A }{c}.
\end{equation}

Physically, inside this surface, the fluid flows inward (despite also rotating), with supersonic radial speed at every point. This means that all acoustic disturbances will be trapped inside this region. Therefore, the acoustic event horizon plays the same role as the horizon of black hole in general relativity.

When the fluid velocity equals to the speed of sound, the $g_{00}$ component of the metric vanishes. This defines the boundary known as the ergosphere
\begin{equation}
  r_e = \frac{\sqrt{A^2 + B^2}}{c}.
\end{equation}
Since $r_e>r_h$, the ergosphere lies outside the horizon. The region between the ergosphere and the horizon is the ergoregion. Within the ergoregion, the fluid moves at supersonic speed, and the $g_{00}$ component of the metric tensor becomes spacelike with positive values,  which resembles the ergoregion of a rotating black hole in general relativity. The ergosphere also sets the static limit of the acoustic black hole. Inside this region, the phonons (quantization of the sound waves) can not remain stationary with respect to an outside observer. This regions lies in the core of the Penrose effect. By employing its properties, we can extract energy from the acoustic black hole through superadiance phenomena or the analogue Penrose process. As demonstrated in Ref.~\cite{Torres2017}, an incident wave can be amplified by the draining vortex and the extra energy is extracted from it. This phenomenon can be considered as an analogue superadiance effect in acoustic black holes. The analogue Penrose process will be discussed in Sec.~\ref{Analogue Penrose}, after we present some key definitions.

%%%%%%%%%%%%%%%%%%%%%%%%%%%%%%%%%%%%%%%%%%%%%%%%%%%%%%
\subsection{Frame dragging effect and surface gravity}

As mentioned earlier, inside the ergoregion, as a consequence of the rotation of the acoustic hole, phonons can not remain stationary. Because of this, particles in the acoustic spacetime will experience an interesting frame dragging effect. In general relativity, a rotating black hole can distort the spacetime metric and leads to the orbit of a nearby test particle precess around it. This is the frame dragging, or Lense-Thirring effect~\cite{Carrollbook}. For the rotating acoustic black hole described by the metric in Eq.~\eqref{metric}, we have a similar frame dragging effect. Inside the ergoregion, since $g_{tt}\, ,g_{rr}\, ,g_{\phi\phi}>0$, any massivi particle moving in a timelike trajectory, $\dd s^2<0$, requires motion in the $\phi$ direction. Let us consider a particle with momentum $\left( p_t, \, p_r, \, p_{\phi} \right)$ moving in this acoustic spacetime. Then
\begin{equation}
    \begin{split}
          p^t & =  g^{t a} p_a = g^{t t} p_t + g^{t \phi} p_{\phi}, \\
  p^{\phi} & =  g^{\phi a} p_a = g^{\phi t} p_t + g^{\phi \phi} p_{\phi}.
    \end{split}
\end{equation}
Hence, because of the mixing components of the metric tensor, even a particle with no initial azimuthal velocity ($p_{\phi}=0$) will get a nonzero angular momentum $p^{\phi}=g^{\phi t}p_t$ under the influence of the acostic black hole. So, inside the ergoregion, a particle will be dragged along the $\phi$ direction regardless of its energy or state of motion.

To more quantitatively characterize the frame dragging effect, we consider the metric given in Eq.~\eqref{metric}, the corresponding metric tensor is given by
\[ g_{\mu \nu} = \left(\begin{array}{ccc}
     - \left( c^2 - \frac{A^2 + B^2}{r^2} \right) & 0 & - B\\
     0 & \left( 1 - \frac{A^2}{c^2 r^2} \right)^{- 1} & 0\\
     - B & 0 & r^2
   \end{array}\right) \]
 The inverse of this metric can be calculated as
\[ g^{\mu \nu} = \left(\begin{array}{ccc}
     \frac{r^2}{A^2 - c^2 r^2} & 0 & \frac{B}{A^2 - c^2 r^2}\\
     0 & 1 - \frac{A^2}{c^2 r^2}  & 0\\
     \frac{B}{A^2 - c^2 r^2} & 0 & \frac{A^2 + B^2 - c^2 r^2}{A^2 r^2 - c^2
     r^4}
   \end{array}\right) \]

Supposing now we drop a  particle into this acoustic black hole from infinity with zero angular momentum. Such particle will acquire an angular velocity in the direction of the rotating acoustic black hole
\begin{equation}
  \frac{d \phi}{d t} = \frac{d \phi / d \tau}{d t / d \tau} =
  \frac{p^{\phi}}{p^t} = \frac{g^{\phi t} p_t + g^{\phi \phi} p_{\phi}}{g^{t
  t} p_t + g^{t \phi} p_{\phi}} = \frac{g^{\phi t}}{g^{t t}} = \frac{B}{r^2}.
\end{equation}
Therefore, a particle coming from infinity with zero angular momentum will pick up an angular velocity from  the effective acoustic spacetime. The frame dragging effect continues as the particle approaches the event horizon at $r_h$, where it becomes
\begin{equation}
  \Omega_h = \frac{B c^2}{A^2}.
\end{equation}
$\Omega_h$ is the minimum angular velocity of a particle at the horizon, we can defined it as the angular velocity of the event horizon. %\textcolor{red}{LCC: This sentence is strange. Angular velocity and angular momentum are two distinct things, so we cannot define one as being the other.}

We can now construct the Killing vector normal to the horizon and compute the surface gravity of the acoustic black hole. The acoustic metric shown in Eq.~\eqref{metric} has no explicit dependence on time coordinate $t$ and on the angular variable $\phi$. This means that the associated acoustic spacetime is invariant under time translation and rotation along the $\phi$ direction. Hence, we can define two Killing vectors for this metric ---one is related to the time-translation invariance, $(k_1)^{\mu}=(1,0,0)$, and the other related to the rotational symmetry, $(k_2)^{\mu}=(0,0,1)$. Because the metric~\eqref{metric} is stationary but not static, and the acoustic black hole is rotating, the acoustic horizon at $r_h$ is not the Killing horizon for the asymptonic time-translation Killing vector $k^{\mu}$.  We can check this by computing the norm of $k^{\mu}$
\begin{equation}\label{knorm}
    (k_1)^{\mu} (k_1)_{\mu} = - c^2 + \frac{A^2 + B^2}{r^2}, 
\end{equation}
which does not vanish at the event horizon. However, as we can see, the norm of the $k^{\mu}$ vanishes at the ergosphere, which indicates the ergosphere as the stationary limit surface.

Since a linear combination of the Killing vector $(k_1)^{\mu}$ and $(k_2)^{\mu}$ is also a Killing vector of this spacetime, we can use the  angular velocity of the event horizon to construct a Killing vector normal to the horizon, $l^{\mu} = (k_1)^{\mu}+\Omega_h (k_2)^{\mu}$, whose norm vanishing at the horizon 
\begin{equation}
    l^{\mu} l_{\mu} = \left[ \frac{(A^2 - c^2 r^2) (A^4 + A^2 B^2 - B^2 c^2
   r^2)}{A^4 r^2} \right]_{r = r_h} = 0.
\end{equation}
The surface gravity of this Killing horizon is defined as
\begin{equation}
    \kappa^2 = \frac{1}{2} \nabla_a l_b \nabla^a l^b,
\end{equation}
resulting in
\begin{equation}\label{surfacegravity}
    \kappa = \frac{c^2}{A }.
\end{equation}
So the surface gravity is a constant at the horizon, being determined by  the sound velocity and the parameter $A$. In general relativity, the surface gravity denotes the acceleration of an observer near the horizon. While for the acoustic black hole, the surface gravity describes the acceleration of the fluid when it crosses the horizon~\cite{Barcelo2011}. Because of the acceleration of phonons around the event horizon, the acoustic event horizon can emit Hawking radiation of thermal phonons with a temperature determined by the surface gravity. This effect resembles the Hawking effect, which motivates  Unruh's analogue gravity research~\cite{Unruh1981}. Moreover, in general relativity, the constant values of surface gravity over the horizon is directly linked to the zeroth law of black hole mechanics~\cite{Padmanabhan2010}. However, because the Einstein equation is absent for the analogue gravity, we can not directly connect Eq.~\eqref{surfacegravity} with the zeroth law of black hole mechanics. We will discuss this issue in the end of the paper.

\subsection{Komar mass and angular momentum}\label{Sec:Komarmass}

Apart from the surface gravity, the mass and angular momentum of black holes also have important influences on its properties and dynamics. Here, we are going to investigate the effective mass and angular momentum of the acoustic black hole. Although we can obtain an effective acoustic metric to simulate some kinetic properties of general relativity, the underline physical systems in the analogue case follow different sets of dynamic equations, and Einstein's field equation is absent. In  general relativity, the curvature of spacetime is determined by the energy-momentum tensor. While in the analogue gravity case, the effective acoustic spacetime is determined by the background velocity field. Without the the Einstein equation, we can not directly link the acoustic metric with underline mass and momentum of the system. However, we can define these effective quantities for the acoustic black hole by resorting on a different approach.

We remember that the effective acoustic metric is pseudo-Riemannian with signature $(-,+,+)$, asymptotically flat and posses time-translational and rotational symmetry. Therefore, according to Noether's theorem, there will be two conserved quantities corresponding to these symmetries. By using the Killing vectors discussed in the previous subsection, we can define the conserved quantities as the effective mass and angular momentm of the acoustic black hole with the Komar integral. In general relativity, for any stationary spacetime, the Komar integral can be used as a formal method to calculate the mass and angular momentum of black holes~\cite{Carrollbook}. For the analogue gravity case,  we expect the Komar mass and angular momentum may be suitable definitions given the involved symmetries. We can identify the conserved quantity corresponding to the time-translational Killing vector $(k_1)^{\mu}$ as the effective masss, while the conserved quantity corresponding to the rotational Killing vector $(k_2)^{\mu}$ should be regarded as the angular momentum. For the metric given in Eq.~\eqref{metric}, using the Killing vectors $(k_1)^{\mu}$ and $(k_2)^{\mu}$, we can defined the corresponding Komar mass $M$ and angular momentum $J$ as~\cite{Cohen1984,Modak2012}
\begin{equation}\label{Komarint}
    \begin{split}
        M  = & - \frac{1}{8 \pi} \int_{\Sigma} \star (\dd\mathbf{k}_1),\\
  J  = & \frac{1}{16 \pi} \int_{\Sigma} \star (\dd\mathbf{k}_2) ,   
    \end{split}
\end{equation}
respectively. The $\star-$ symbol denotes the Hodge dual~\cite{Carrollbook}

With the time-translational Killing vector, the exterior derivative of $(k_1)_{\mu}$ can be straightforwardly calculated as
\begin{equation}
     \dd k_1 = \frac{2 (A^2 + B^2)}{r^3} \dd t \wedge \dd r 
\end{equation}
The corresponding Hodge dual is then given by
\begin{equation}
    \star (\dd k_1) = \frac{2 (A^2 B + B^3)}{r^4 c} \dd t - \frac{2 (A^2 + B^2)}{r^2
   c} \dd \phi 
\end{equation}
Since the mass of the acoustic black hole should be defined for simultaneous event, the $\dd t$ term should be neglected for the Komar integral, which lead to the effective Komar mass on the horizon
\begin{equation}\label{KMass}
    M = \frac{1}{8 \pi} \int_H \frac{2 (A^2 + B^2)}{r^2 c} d \phi = \frac{(A^2 +
   B^2)c}{2 A^2 }.
\end{equation}
As  we can see, since the acoustic metric is determined by the velocity profile, the effective mass depends on the velocity parameters $A$ and $B$. As anticipated, unlike the general relativity case the effective mass of the acoustic metric is not linked with the energy and momentum of the underlying physical system (the fluid).

In the same way, the exterior derivative of the rotational Killing vector $(k_2)_{\mu}$ can be calculated as
\begin{equation}
    \dd k_2 = 2 r \dd r \wedge \dd \phi,
\end{equation}
while the corresponding Hodge dual is given by
\begin{equation}
   \star (\dd k_2) = \frac{2 (c^2 r^2 - A^2 - B^2)}{r^2 c} \dd t + \frac{2 B}{c} \dd
   \phi .
\end{equation}
By using the Komar definition in Eq.~\eqref{Komarint},  the angular momentum of the acoustic black hole is given by
\begin{equation}
    J = \frac{1}{16 \pi} \int_H \frac{2 B}{c} \dd \phi = \frac{B}{4 c}.
\end{equation}
So the effective angular momentum of the acoustic metric is determined by the azimuthal velocity parameter $B$, which makes it directly linked with the underline angular momentum of the vortex. If the vortex is quantized, as in superfliud helium or Bose-Einstein condensates~\cite{Dalfovo1999,Pethick2008}, then the angular momentum of the vortex can only take 
discrete values and the parameter $B$ will be proportional to an integer number. In this case, the Komar angular momentum and mass of the acoustic black hole are also discrete. This arises from the quantization of the circulation flow in underline physical system, which leads to the interesting phenomena of mass and angular momentum quantization of acoustic black hole.

%%%%%%%%%%%%%%%%%%%%%%%%%%%%%%%%%%%%%%%%%%%%%%%%
%%%%%%%%%%%%%%%%%%%%%%%%%%%%%%%%%%%%%%%%%%%%%%%%
%%%%%%%%%%%%%%%%%%%%%%%%%%%%%%%%%%%%%%%%%%%%%%%%
\section{Analogue Penrose Process}\label{Analogue Penrose}

An interesting  property of the rotating black hole in general relativity is the possibility of extracting energy from it through the Penrose process~\cite{Penrose1971}. Since the acoustic spacetime arising from the draining bathtub model can simulate many properties of the rotating black hole, including the ergoregion and the frame dragging effect, we expect to also observe the analogue Penrose process in this acoustic spacetime.

To simulate the Penrose process, we can send a massive particle travelling along a geodesic trajectory into the black hole. The momentum of the particle is given by
\begin{equation} 
p^{\mu} = m \dv{x^{\mu}}{\tau} ,
\end{equation}
where $m$ is the rest mass of the particle. Because the particle travels along a geodesic, we can define its energy and angular momentum as
\begin{equation}\label{ELdefinition}
\begin{split}
  E  = & - (k_1)_u p^{\mu} = B m \dv{\phi}{\tau} + m \left( c^2 - \frac{A^2
  + B^2}{r^2} \right) \dv{t}{\tau}, \\
  L  = & (k_2)_{\mu} p^{\mu} = r^2 m \dv{\phi}{\tau} - B m \dv{t}{\tau} ,
\end{split}
\end{equation}
respectively. 

We have inserted a negative sign in the definition of the energy to make it positive. This is because outside the ergoregion, the Killing vector and the momentum $p^{\mu}$ are both timelike, so their inner product is negative. Hence, under this definition, along the geodesic, the particle energy is conserved and remains positive even when the particle enters the ergoregion.

Once the particle enters into the ergosphere, it breaks into two parts with momenta $p^{(1) \mu}$ and $p^{(2) \mu}$, respectively. The conservation of four-momentum requires that
\begin{equation}\label{momenconser}
  p^{(0) \mu} = p^{(1) \mu} + p^{(2) \mu} .
\end{equation}
Contracting it with the Killing vector,  we have the energy conservation
\begin{equation}\label{Energy}
    E^{(0)} = E^{(1)} + E^{(2)}.
\end{equation}
Inside the ergoregion, the norm of the Killing vector in Eq.~\eqref{knorm} becomes positive, so the Killing vector $(k_1)^{\mu}$ becomes spacelike. According to the energy definition in Eq.~\eqref{ELdefinition}, a particle can have negative energy inside the ergosphere. Since all particles must have positive energies, if they are outside the ergoregion, a particle inside the ergosphere with negative energy must either remains in the ergosphere, or be accelerated until its energy is positive in order to escape to infinity.

The analogue Penrose process follows by picking up the second  particle  with negative energy $E^{(2)}<0$ inside the ergoregion, then the first particle must have positive energy $E^{(1)}>0$. Each of these smaller particles then follows their own geodesics. Because $E^{(2)}<0$, the second particle can not escape the ergoregion, and typically it will fall into the acoustic black hole. Then, by the energy relation Eq.~\eqref{Energy}, the  first particle will have energy bigger than $E^{(0)}$
\begin{equation}
  E^{(1)} = E^{(0)} - E^{(2)} > E^{(0)}.
\end{equation}
So we expect the first particle will  escape from the ergosphere with more energy than the initial particle energy $E^{(0)}$. Thus by using the existence of negative energy orbits in the ergoregion and the local conservation of energy for process taking place inside the ergoregion, we can extract energy from the acoustic black hole. This is known as the Penrose effect.

In the analogue Penrose process, the extra energy extracted from the acoustic black hole comes from the decrease of its mass and angular momentum. Since the combined Killing vector $l$ is future-pointing, we can use it to find a restriction on the amount of energy that can be extracted by this process. For the second particle with $E^{(2)}<0$ falling into the acoustic black hole, the inner product of its momentum and the Killing vector must be negative,
\begin{equation}
p^{(2) \mu} l_{\mu} = - E^{(2)} + \Omega_h L^{(2)} < 0,
\end{equation}
which leads to the inequality
\begin{equation}
  L^{(2)} < \frac{E^{(2)}}{\Omega_h} . \label{Jmin}
\end{equation}
Since $E^{(2)} $ is negative and $\Omega_h$ is positive, so the angular momentum of the second particle, $L^{(2)}$, must be negative, which means the particle moves against the rotation of the acoustic black hole.

Once the first particle with positive energy has escaped from the ergosphere, and the second particle with negative energy has fallen inside the horizon, the mass and angular momentum of the acoustic black hole will be changed by the negative contributions of the second particle
\begin{eqnarray*}
  \delta M & = & E^{(2)},\\
  \delta J & = & L^{(2)} .
\end{eqnarray*}
Along with Eq.~\eqref{Jmin}, we can set a limit on the change of the angular momentum of the acoustic black hole
\begin{equation}\label{angneq}
    \delta J < \frac{\delta M}{\Omega_h} . 
\end{equation}
%To reach this limit, the second particle should become more and more null \textcolor{red}{LCC: WHAT IS THE MEANING OF A PARTICLE BECOMING NULL?}, which results in the ideal analogue Penrose process with $\delta J=\delta M/\Omega_h$. 
During this process, the variations of the effective mass and angular momentum of the acoustic black hole are governed by the dynamical equations of the system. In contrast to the general relativity, in which the variations of the mass and angular momentum of rotating black holes are determined by the Einstein's equation. This is the most important fundamental difference between the analogue gravity and general relativity.

As we can see, the analogue Penrose process relies on the conservation of momentum and the energy is extracted from the rotation of the acoustic black hole. From the effective mass and angular momentum of the acoustic black hole defined in Sec.~\ref{DBT metric}, the angular  momentum of the acoustic black is determined by the  parameter $B$. Although vortices can be regarded as stable objects, the circulation of vortices can be changed with various methods. For classical fluid with viscosity, the circulation of the vortex can be reduced through a dissipating process~\cite{Saffman1993}.  In a quantum fluid, the circulation of a vortex is quantized with a winding number, which makes the quantum vortices  topologically stable~\cite{Fischer1999}. But since the kinetic energy of the quantum vortex is proportional to the square of winding number, the quantum vortex with high winding number is metastable and it can break into mutli-vortices configuration with lower energy~\cite{Pethick2008}. As demonstrated in the superradiance experiment, the rotating energy of the vortex can be extracted with directed waves, and the circulation of vortex is reduced  during the process~\cite{Torres2017}. For the quantum vortex, its winding number and circulation can also be reduced when combined with antivorteces~\cite{Solnyshkov2019}. In our work, during the analogue Penrose process, the angular momentum and the mass of the acoustic black hole are reduced. Since the rotation of the acoustic black hole is determined by the circulation parameter $B$, we expect it will change during the analogue Penrose process. While the parameter $A$ is linked with the horizon radius and has no connection with the angular momentum, so the parameter $A$ will remain the same in the process. Hence, during the analogue Penrose process, we can fix the parameter $A$ to be a constant, the mass of the acoustic black hole varies with the parameter $B$ as
\begin{equation}
    \delta M = \frac{cB\delta B}{A^2}.
\end{equation}
While  the variation of the acoustic black hole angular momentum is given by
\begin{equation}
    \delta J = \frac{\delta B}{4c}.
\end{equation}
Hence, we can check that the inequality~\eqref{angneq} indeed holds for the rotating acoustic metric given in Eq.~\eqref{metric}
\begin{equation}
    \frac{\delta M}{\Omega_h}=\frac{\delta B}{c}>\delta J.
\end{equation}
 Hence, the amount of  angular momentum of the acoustic black hole that can be reduced in the analogue Penrose process is indeed bounded.

To implement the analogue Penrose process, we can use irrotational vortices as the particles to extract energy from the acoustic black hole. However, because it is not easy to break one the vortex into two vortices with opposite angular momentum, here we consider another way to implement the analogue Penrose process with vortices. As demonstrated in Ref.~\cite{Solnyshkov2019}, by creating a vortex-antivortex pair inside the ergoregion with strong localized potential pulses, it is possible to simulated the analogue Penrose process. During the process, because of the attractive interaction between the acoustic black hole and the antivortex, the antivortex falls into the effective acoustic black created by a dissipating giant vortex with high winding number, while the vortex escapes from the ergosphere with more kinetc enenrgy. Following the same spirit, here, we can simulate the analogue Penrose process with the vortex-antivortex pair in the ergospher region. 

To simulate the analogue Penrose process, we choose a small irrotational vortex and an antivortex as our operating particles to extract energy from the acoustic black hole. The circulations of this irrotational vortex is given by $\Gamma=\oint \mathbf{v}\cdot d\mathbf{r}=2\pi\gamma$, where $\gamma$ is the circulation parameter, which is  smaller than the circulation parameter of the acoustic black hole~\cite{Fischer2002}. For the irrotational antivortex, the magnitude of its circulation is the same, but it is moving along an opposite direction. Hence, we can denote the circulation parameters of the small vortex and antivortex as $\pm\gamma$, respectively. For the small irrotational vortex, its velocity only contains the azimuthal component. In the acoustic spacetime, the effective acoustic metric of such vortex 
resembles a cosmic string~\cite{Fischer2002}. By using the Komar integral, we can define the effective mass $m_v$ and angular momentum $j_v$ of the small irrotational vortex as $m_v = \gamma^2c/(2A^2)$ and $j_v=\gamma/(4c)$ (the detailed derivation is given in appendix~\ref{appendix1}). The masses $m_{\bar{v}}$ of both the vortex and the antivortex (the ones representing the particles) are the same, but their angular momenta $j_{\bar{v}}$ are opposite, which makes the vortex and antivortex behave like a particle-antiparticle pair. Now, to simulate the analogue Penrose process, we can create a vortex-antivortex pair in the ergoregion. As demonstrated in Ref~\cite{Solnyshkov2019}, the vortex-antivortex pairs can be created by a strong Gaussian-shaped potential pulse in the polariton BEC. Once the vortex-antivortex pair is created, the vortex and the antivortex will behave like massive particles in the acoustic spacetime. Since the vortex and antivortex behave like anti-particles, following the conservation of momentum, the summation of the momenta of the vortex and antivortex should be zero,
\begin{equation}
    p_v + p_a = 0.
\end{equation}
Hence, we must also have the conservation of energy 
\begin{equation} 
    E_v + E_a = 0.
\end{equation}
Because the vortex is rotating along the same direction as the central vortex, we can define its energy to be positive. While the anti-vortex is counter-rotating with respect to the acoustic black hole, its energy will be negative. Now, because of the attractive interaction exerted by the acoustic black hole, the antivortex will drop into it. When the anti-vortex has crossed the event horizon, the angular momentum of the acoustic black hole will decrease by the amount $j_{\bar{v}}$. Also, its mass will be reduced by the amount of $m_{\bar{v}}$. The vortex with positive energy will experience an repulsive interaction exerted by the central vortex, consequently accelerating and escaping from the the ergosphere. When this happens, it will have more kinetic energy than it had in the beginning. So in this analogue process, by using the vortex-antivortex pair to simulate particles, we can reduce the mass and the angular momentum of the acoustic black hole and increase the kinetic energy of the escaping vortex. This is the Penrose effect.

In the above discussion, we have also defined the effective mass and angular momentum of the operating vortices with Komar integral. These definitions are different from the effective mass defined with the kinetic energy of the vortex, as considered in Ref.~\cite{Solnyshkov2019}, where the topologically protected vortex is regarded as an elementary object composed by a collection of atoms~\cite{Fischer1999}. Since we are analysing the Penrose process in the acoustic spacetime, and the vortices behave like cosmic strings in this spacetime, we believe that the Komar mass and angular momentum should be used as a suitable definitions for the vortices. In addition, with these effective mass and angular momentum, we can compare them with the mass and angular momentum of the acoustic black hole (the central vortex), in order to obtain a consistent theoretical analysis of the analogue Penrose process.

When the parameter $B$ is reduced to zero, the angular momentum of the acoustic black hole becomes zero. In this limit, we can no longer extract energy from the black hole by the Penrose process. So the maximum amount of energy can be extracted from the acoustic black is given by
\begin{equation}
    \Delta E =  \frac{cB^2}{2A^2}.
\end{equation}
In this case, the Komar mass of the acoustic black hole turns into
\begin{equation}
    M_i=\frac{cA^2}{2A^2}=\frac{c}{2}.
\end{equation}
This is the smallest mass of the acoustic black hole, which can also be regarded as the irreducible mass of the acoustic black hole. In the analogue Penrose process, the acoustic black hole exerts an attractive interaction on the antivortex, which leads the antivortex to fall into the black hole. When the antivortex falls into the acoustic black hole, it also exert influence on the acoustic back, the angular momentum and mass of the acoustic black hole are reduced, and the effective acoustic spacetime is also modified by the antivortex. This feedback effects resemble the typical backreaction effect in general relativity. In general relativity, the backreaction effect experienced by the spacetime is determined by the Einstein's equation. However, in this acoustic spacetime, this backreaction effect is governed by the underline dynamical equations of the system.

Furthermore, as an analogue of the rotating black hole in general relativity, we can also define the effective area of the event horizon. With the metric given in Eq.~\eqref{metric}, by setting $r = r_h$, $\dd t = 0$ and $\dd r = 0$, then the metric on the event horizon is given by
\begin{equation}
    \gamma_{i j} \dd x^i \dd x^j = \dd s^2 (\dd t = 0, \dd r = 0, r = r_h) = r_h^2 \dd\phi^2.
\end{equation}
The horizon area is the integral of the induced area element
\begin{equation}
    S = \int \sqrt{\gamma} \dd \phi = \int r_h \dd \phi = 2 \pi r_h = \frac{2 \pi
   A}{c}.
\end{equation}
So the horizon area of the acoustic black hole has no dependence on the parameter $B$. During the analog Penrose process, the change of the angular momentum will not lead to variance of the horizon area, which is in striking contrast to the real black holes predicted by general relativity and state the limitations of the analogue model, which lacks Einstein's field equations.

Another important aspect of the analogue Penrose process to be noted, in quantum superfluid (such as superfliud helium, Bose-Einstein condensate),  the circulation of the vortex is quantized and proportional to an integer winding number~\cite{Dalfovo1999,Pethick2008}. In this case, the angular momentum and mass of the acoustic black hole and the irrotational vortex  will also be quantized. So in the analogue Penrose process, the mass and angular momentum of the acoustic black hole will be reduced by  discrete amounts, but the reduced angular momentum and mass of the acoustic black are still equal  to the reduced angular momentum and mass of the antivortex.

For the possible experimental realization of the analog Penrose process, the polariton BEC  and the shallow water tank platform can be used. Because the pumping-dissipating nature of polariton BEC, the dissipating vortex can be created for simulating the rotating acoustic spacetime ~\cite{Solnyshkov2019}. As shown in Ref~\cite{Jacquet2020}, with focused Laguerre–Gauss beam, the circulation of the dissipating vortex can also be tuned, which makes the Polariton fluids system as ideal platform to simulate the analogue Penrose process.  While for the classical water tanker, the superradiance phenomenon have been experimental realized with similar dissipating vortex\cite{Torres2017} . The analog Penrose process can also tested with the similar experiment setup. Since the vortex angular momentum would be quantized in the polariton BEC, the angular momentum of the acoustic black hole will be decreased by an amount propotioanl to an integer number in the analog Penrose process. In the water tanker setup, because the system is classical fluid, there is no quantization of vortex angular momentum  and the angular momentum of the acoustic black hole can be reduced continuously.

%%%%%%%%%%%%%%%%%%%%%%%%%%%%%%%%%%%%%%%%%%%%%%%%%%%%%%%%%
%%%%%%%%%%%%%%%%%%%%%%%%%%%%%%%%%%%%%%%%%%%%%%%%%%%%%%%%%
%%%%%%%%%%%%%%%%%%%%%%%%%%%%%%%%%%%%%%%%%%%%%%%%%%%%%%%%%
%\subsection{Thermodynamics of the acoustic black hole}

%\textcolor{red}{LCC: I REALLY DO NOT KNOW IF THIS SECTION IS NECESSARY. IT SEEMS TO ME THAT THE COMMENTS HERE DO NOT ADD ANY NEW PHYSICS TO THE PROBLEM.}

%%%%%%%%%%%%%%%%%%%%%%%%%%%%%%%%%%%%%%%%%%%%%%%%%%%%%%%%
\section{Conclusion and Discussion}\label{conclusion}

We have studied the analog Penrose process in a rotating acoustic black hole based on the draining bathtub model. Because of the rotating nature of this acoustic spacetime, a particle, which is represented here by a vortex moving on the quantum fluid, will experience the frame dragging effect. The Penrose process is here considered in terms of the Komar mass and angular momentum, which allow us to characterize the extracted energy in terms of the physical quantities of the analogue spacetime. However, because in the analog gravity setting we do not have the Einstein's equations to link the acoustic spacetime with the underline physical energy-momentum tensor, the black hole thermodynamic laws can not be established in this case. This is the typical limitation of the analogue gravity research, where distinct sets of equations govern the system and the spacetime. 

Another interesting point to note, with the effective mass $M$, the angular momentum $J$, and the surface gravity at the horizon $\kappa$, the Komar mass can be expressed is terms of the area $S$ of the horizon and of the angular momentum
\begin{equation}
    M = \frac{\kappa S}{4 \pi} + 2 \Omega_h J.
\end{equation}
This relation is very similar to the one we obtain in a rotating black hole in general relativity. But because the Einstein equation is absent for the analog models of gravity, we can not directly link the variation of $M$ and $J$ with the first law of black hole thermodynamics.

%%%%%%%%%%%%%%%%%%%%%%%%%%%%%%%%%%%%%%%%%%%%%%%
\section{Acknowledgments}

We acknowledge National Natural Science Foundation of China grant (NSFC) (12075145), Shanghai Government grant STCSM (2019SHZDZX01-ZX04). LCC acknowledges financial support from the Brazilian agencies CNPq (PQ Grant No. 305740/2016-4 and INCT-IQ 246569/2014-0) and FAPEG (PRONEX 201710267000503).

%%%%%%%%%%%%%%%%%%%%%%%%%%%%%%%%%%%%%%
%
%		APPENDICES
%
%%%%%%%%%%%%%%%%%%%%%%%%%%%%%%%%%%%%%%
\appendix

\section{The mass and angular momentum of the irrotational operating vortex}\label{appendix1}

To implement and test the analog Penrose process, we need to drop a massive quasi-particle into  the ergosphere region. Here, we consider the irrotational vortex. For an irrotational vortex, the circulation around the vortex is $\Gamma=\oint \mathbf{v}\cdot d\mathbf{r}=2\pi\gamma$. The velocity profile of the irrotational vortex is determined by the circulation~\cite{Fischer2002}
\begin{equation}\label{vvortex}
    \mathbf{v}=\frac{\Gamma}{2\pi r}\hat{\phi} = \frac{\gamma}{r}\hat{\phi} .
\end{equation}
In contrast to the dissipating vortex of draining bathtub model, the irrotational vortex don't have a radial flow. Applying the acoustic spacetime approach to the irrotational vortex, we can obtain the effective acoustic metric of the irrotational vortex in $2+1$ dimension as
\begin{equation}
    g_{\mu\nu}^{v} =  \left(\begin{array}{ccc}
     - \left( c^2 - \frac{\gamma^2}{r^2} \right) & 0 & - \gamma\\
     0 & 1 & 0\\
     - \gamma & 0 & r^2
   \end{array}\right) .
\end{equation}
This metric resembles the cosmic strings in Lorentz spacetime,  and at large $r$, it agrees with the metric of a massless spinning cosmic string asymptotically\cite{Fischer2002}. Futhermore, without the radial flow, there is no acoustic event horizon for this metric. But when the flow goes supersonic, the $g^v_{tt}$ component of the metric becomes zeros, which lead to a stationary limit surface locating at
\begin{equation}
    r_s = \frac{|\gamma|}{c}.
\end{equation}
Inside this surface, no observeer can remain stationary.
Because of the symmetry of this metric, we can also find two Killing vectors, which correspond to the time-translation invarance, $(k_1^v)^{\mu}=(1,0,0)$ and the  the rotational symmetry, $(k^v_2)^{\mu}=(0,0,1)$. By applying the Komar definition to the irrotational vortex metric, we can define the effective mass and angular momentum of the irrotational vortex in the acoustic spacetime as
\begin{equation}\label{Komarvortex}
    \begin{split}
        m_v  = & - \frac{1}{8 \pi} \int_{\Sigma} \star (d\mathbf{k}_1^v)=\frac{1}{8\pi}\int_{r_h}\frac{2\gamma^2}{r^2c}d\phi=\frac{\gamma^2}{2r_h^2c}=\frac{\gamma^2 c} {2A^2},\\
  j_v  = & \frac{1}{16 \pi} \int_{\Sigma} \star (d\mathbf{k}_2^v)=\frac{1}{16\pi}\int_{r_h}\frac{2\gamma}{c}d \phi=\frac{\gamma}{4c} ,   
    \end{split}
\end{equation}
where, we have defined the effective mass and angular momentum of the irrotational vortex at the horizon of the acoustic black hole. As we can see, the vortex living in the acoustic spacetime is like a massive particle, with its mass and angular momentum determined by the circulation parameter.


\begin{thebibliography}{99}

\bibitem{Gordon1923} W. Gordon. Zur Lichtfortpflanzung nach der Relativit\"{a}tstheorie. Ann. Phys. \textbf{72}, 421 (1923).

\bibitem{Felice1971} F. de Felice, On the gravitational field acting as an optical medium. Gen. Relativ. Gravit. \textbf{2}, 347 (1971).

\bibitem{Unruh1981}  W. G. Unruh. Experimental black hole evaporation? Phys. Rev. Lett. \textbf{46}, 1351 (1981).

\bibitem{Lahav2010} O. Lahav, A. Itah, A. Blumkin, C. Gordon, S. Rinott, A. Zayats, and J. Steinhauer. Realization of a sonic black hole analogue in a Bose-Einstein condensate. Phys. Rev. Lett. \textbf{105}, 240401 (2010).

\bibitem{Garay2000}  L. J. Garay, J. R. Anglin, J. I. Cirac, and P. Zoller. Sonic analogue of gravitational black holes in Bose-Einstein condensates. Phys. Rev. Lett. \textbf{85}, 4643 (2000).

\bibitem{Yatsuta2020} I. Yatsuta, B. Malomed, and A. Yakimenko. Acoustic analogue of Hawking radiation in quantized circular superflows of Bose-Einstein Condensates. Phys. Rev. Res. {\bf 2}, 043065 (2020).

\bibitem{Nova2019} J. R. Mu\~noz de Nova, K. Golubkov, V. I. Kolobov, and J. Steinhauer. Observation of thermal Hawking radiation and its temperature in an analogue black hole. Nature \textbf{569}, 688 (2019).

\bibitem{Steinhauer2014} J. Steinhauer. Observation of self-amplifying Hawking radiation in an analogue black-hole laser. Nat. Phys. \textbf{10}, 864 (2014).

\bibitem{Kolobov2021} V. I. Kolobov, K. Golubkov, J. R. Mu\~noz de Nova, and J. Steinhauer. Observation of stationary spontaneous Hawking radiation and the time evolution of an analogue black hole. Nat. Phys. \textbf{17}, 362 (2021).

\bibitem{Steinhauer2017} J. Steinhauer and J. R. M. De Nova. Self-amplifying Hawking radiation and Its background: A numerical study. Phys. Rev. A \textbf{95}, 1 (2017).

\bibitem{Eckel2018} S. Eckel, A. Kumar, T. Jacobson, I. B. Spielman, and G. K. Campbell. A rapidly expanding Bose-Einstein condensate: An expanding universe in the lab. Phys. Rev. X \textbf{8}, 21021 (2018).

\bibitem{Weinfurtner2011} S. Weinfurtner, E. W. Tedford, M. C. J. Penrice, W. G. Unruh, and G. A. Lawrence. Measurement of stimulated Hawking emission in an analogue system. Phys. Rev. Lett. \textbf{106}, 021302 (2011).

\bibitem{Torres2017} T. Torres, S. Patrick, A. Coutant, M. Richartz, E. W. Tedford, and S. Weinfurtner. Rotational superradiant scattering in a vortex flow. Nat. Phys. \textbf{13}, 833 (2017).

\bibitem{sheng2013} C. Sheng, H. Liu, Y. Wang, S. N. Zhu, and D. A. Genov. Trapping light by mimicking gravitational lensing. Nat. Photon. \textbf{7}, 902 (2013).

\bibitem{Tinguely2020} R. A. Tinguely and A. P. Turner. Optical analogues to the equatorial Kerr–Newman black hole. Commun. Phys. \textbf{3}, 120 (2020).

\bibitem{Bekenstein2017} R. Bekenstein, Y. Kabessa, Y. Sharabi, O. Tal, N. Engheta, G.Eisenstein, A. J. Agranat, and M. Segev. Control of light by curved space in nanophotonic structures. Nat. Photon. \textbf{11}, 664 (2017).

\bibitem{Drori2019} J. Drori, Y. Rosenberg, D. Bermudez, Y. Silberberg, and U. Leonhardt. Observation of stimulated Hawking radiation in an optical analogue. Phys. Rev. Lett. \textbf{122}, 010404 (2019).

\bibitem{Sheng2018} C. Sheng, H. Liu, H. Chen, and S. Zhu. Definite photon deflections of topological defects in metasurfaces and symmetry breaking phase transitions with material loss. Nat. Commun. \textbf{9}, 4271 (2018).

\bibitem{Zhong2018} F. Zhong, J. Li, H. Liu, and S. Zhu. Controlling surface plasmons through covariant transformation of the spin dependent geometric phase between curved metamaterials. Phys. Rev. Lett. \textbf{120}, 243901 (2018).

\bibitem{Molina2017} A. Rold\'{a}n-Molina, A. S. Nunez, and R. A. Duine. Magnonic black holes. Phys. Rev. Lett. \textbf{118}, 061301 (2017).

\bibitem{Barcelo2011} C. Barcel\'{o}, S. Liberati, and M. Visser. Analogue Gravity, Living Rev. Relativ. {\bf 14}, 3 (2011).

\bibitem{Barcelo2019} C. Barcel\'{o}. Analogue black-hole horizons. Nat. Phys. \textbf{15}, 210 (2019).

\bibitem{Penrose1971} R. Penrose and R. M. Floyd, Extraction of Rotational Energy from a Black Hole, Nature Physical Science \textbf{229}, 177 (1971).


  \bibitem{Solnyshkov2019} D. D. Solnyshkov, C. Leblanc, S. V. Koniakhin, O. Bleu, and G. Malpuech, \textit{Quantum Analogue of a Kerr Black Hole and the Penrose Effect in a Bose-Einstein Condensate}, Phys. Rev. B \textbf{99}, 214511 (2019).
  

  
\bibitem{Visser1998} M. Visser, \textit{Acoustic Black Holes: Horizons, Ergospheres and Hawking Radiation}, Class. Quantum Gravity \textbf{15}, 1767 (1998).

\bibitem{Vocke2018} D. Vocke, C. Maitland, A. Prain, K. E. Wilson, F. Biancalana, E. M. Wright, F. Marino, and D. Faccio, Rotating Black Hole Geometries in a Two-Dimensional Photon Superfluid, Optica \textbf{5}, 1099 (2018).


\bibitem{Basak2003} S. Basak and P. Majumdar, Superresonance from a Rotating Acoustic Black Hole, Class. Quantum Grav. \textbf{20}, 3907 (2003).

\bibitem{Patrick2018} S. Patrick, A. Coutant, M. Richartz, and S. Weinfurtner, Black Hole Quasibound States from a Draining Bathtub Vortex Flow, Phys. Rev. Lett. \textbf{121}, 061101 (2018).

\bibitem{Oliveira2010} E. S. Oliveira, S. R. Dolan, and L. C. B. Crispino, Absorption of Planar Waves in a Draining Bathtub, Phys. Rev. D \textbf{81}, 124013 (2010).

\bibitem{Pethick2008} C. Pethick and H. Smith, Bose-Einstein Condensation in Dilute Gases, 2nd ed (Cambridge University Press, Cambridge, New York, 2008).


\bibitem{Padmanabhan2010} T. Padmanabhan, Gravitation: Foundations and Frontiers (Cambridge University Press, Cambridge, 2010).


  \bibitem{Modak2012}S. K. Modak and S. Samanta, \textit{Effective Values of Komar Conserved Quantities and Their
  Applications}, Int J Theor Phys \textbf{51}, 1416 (2012).


\bibitem{Cohen1984} J. M. Cohen and F. de Felice, The Total Effective Mass of the Kerr–Newman Metric, J. Math. Phys. \textbf{25}, 992 (1984).

\bibitem{Carrollbook} S. M. Carroll, Spacetime and Geometry: An Introduction to General Relativity (Cambridge University Press, Cambridge, 2019).

\bibitem{Dalfovo1999} F. Dalfovo, S. Giorgini, L. P. Pitaevskii, and S. Stringari, Theory of Bose-Einstein Condensation in Trapped Gases, Rev. Mod. Phys. \textbf{71}, 463 (1999).






\bibitem{Saffman1993} P. G. Saffman, Vortex Dynamics, 1st ed. (Cambridge University Press, 1993).

\bibitem{Fischer1999} U. R. Fischer, Motion of Quantized Vortices as Elementary Objects, Annals of Physics \textbf{278}, 62 (1999).
\bibitem{Jacquet2020} M. J. Jacquet, T. Boulier, F. Claude, A. Maître, E. Cancellieri, C. Adrados, A. Amo, S. Pigeon, Q. Glorieux, A. Bramati, and E. Giacobino, Polariton Fluids for Analogue Gravity Physics, Phil. Trans. R. Soc. A. \textbf{378}, 20190225 (2020).


\bibitem{Fischer2002} U. R. Fischer and M. Visser, Riemannian Geometry of Irrotational Vortex Acoustics, Phys. Rev. Lett. \textbf{88}, 110201 (2002).

\end{thebibliography}
\end{document}